\documentclass[a4paper]{jpconf}
\usepackage{graphicx}
\usepackage{mathrsfs}
\usepackage{graphicx}
\usepackage{amsmath}
\graphicspath{{Figures/}}
\usepackage{siunitx}
\usepackage{multirow}

\begin{document}
\title{Experimental Validation for Distributed Control of Energy Hubs}

\author{Varsha Behrunani$^{1,2}$, Philipp Heer$^{2}$ and John Lygeros$^{1}$}

\address{$^{1}$Automatic Control Laboratory, ETH Zurich, Switzerland}
\address{$^{2}$Urban Energy Systems Laboratory, Empa, Dübendorf, Switzerland}

\ead{bvarsha@ethz.ch}

\begin{abstract}
As future energy systems become more decentralised due to the integration of renewable energy resources and storage technologies, several autonomous energy management and peer-to-peer trading mechanisms have been recently proposed for the operation of energy hub networks based on optimization and game theory. However, most of these strategies have been tested either only in simulated environments or small prosumer units as opposed to larger energy hubs. This simulation reality gap has hindered large-scale implementation and practical application of these method.  In this paper, we aim to experimentally validate the performance of a novel multi-horizon distributed model predictive controller for an energy hub network by implementing the controller on a complete network of hubs comprising of a real energy hub inter-faced with multiple virtual hubs. The experiments are done using two different network topologies and the controller shows promising results in both setups. 
\end{abstract}

\section{Introduction}

The increase of renewable energy resources , generation, and efficient conversion and storage technologies in the energy system owing to the rise in energy demand and growing climate concerns has resulted in multi-energy hubs. Therefore, they are gaining relevance as promising sustainable solutions for future energy networks, transforming the system from a conventional grid-centric to a decentralized prosumer-centric model. The optimal operation of energy hubs and peer-to-peer trading increases energy efficiency, flexibility, and eases integration of renewable resources~\cite{MOHAMMADI201833}. However, the resulting decentralisation of decision-making gives rise to challenges in central control and joint optimization of the network both in terms of computational tractability and privacy. To mitigate these problems, a novel multi-horizon distributed model predictive control (MH-DMPC) method has been developed for a network of energy hubs and shown to perform remarkably well in simulation~\cite{behrunani_journal}. More generally, while several recent works have covered the area of energy hub control~\cite{AR_smith}, most techniques have not been applied to existing energy hub networks. This leaves a large gap between the theoretical techniques and the practical implementations. The goal of this work is to close this simulation-reality gap by implementing the proposed MH-DMPC on an existing energy hub system in real time to prove its viability, and validate the performance of the control strategy.

Our experiment campaign is based on the NEST building, a district and energy hub demonstrator at Empa in Dübendorf, Switzerland, that is designed to test new technologies and hosts a wide variety of components that convert and store energy~\cite{NEST}. In this paper, the NEST building is interfaced with virtual energy hubs creating a hybrid physical-simulation environment to test the performance of the controller within a network of hubs. This setup demonstrates how the distributed controller designed for multiple energy hubs can be experimentally validated even using just a single real hub by leveraging the simulated environment. Two experiments are conducted, first using a network of hubs determined in simulation and second, using a network comprising of real buildings in the vicinity of the NEST at the Empa campus. The paper is structured as follows: In Section 2, we describe the control methodology. The experimental setup and case study are elaborated in Section 3 and the results are presented in Section 4.



\section{Control Methodology}
\subsection{Energy hub Optimization}
Consider a network of $N$ energy hubs. Each hub comprises different generation, conversion, and storage devices and can import from the electrical and gas grid in order to fulfil its respective electricity and thermal demand. The hubs can also sell surplus electricity to the electricity grid as well as trade electrical energy and thermal energy with other hubs. The goal of our optimal energy hub control is to minimize the total energy costs of the network. The optimal scheduling of the energy hubs is formulated as a finite-horizon economic dispatch problem:
\begin{subequations}
\label{main_opt}
\begin{align} \label{ehub_cost}
\underset{u_{\text{i}}, u_{\text{tr}}, x_{\text{i}}, \epsilon_{\text{i}}}{\operatorname{minimize}} & \ \sum_{i=1}^{N} \sum_{k=0}^{T-1}  J^{k}_{\text{i}} \left(x^{k}_{\text{i}}, u^{k}_{\text{i}}, u^{k}_{\text{tr}}\right) + \lambda_{\epsilon} \| \epsilon^{k}_{\text{i}} \|_2^2  \\ \label{ehub_dyn}
\text{subject to  } & x^{k+1}_{\text{i}}=f_{\text{i}}\left(x^{k}_{\text{i}}, u^{k}_{\text{i}}\right)  , \\ \label{ehub_eq}
& g_{\text{i}}\left(x^{k}_{\text{i}}, u^{k}_{\text{i}}\right) \leq 0 , \ \ \ \ \ \ \quad h_{\text{i}}\left(x^{k}_{\text{i}}, u^{k}_{\text{i}} \right) = 0  ,\\ \label{ehub_loadbalance1}
&l^{k}_{\text{e,i}} = e_{\text{i}}(u^{k}_{\text{i}}) + r^{\text{T}}_{\text{i}} u^{k}_{\text{tr}} , \ \ \ l^{k}_{\text{h,i}} = h_{\text{i}}(u^{k}_{\text{i}}) + \epsilon^{k}_{\text{i}} + s^{\text{T}}_{\text{i}} u^{k}_{\text{tr}}  ,\\ \label{ehub_tr1}
& g_{\text{tr}}\left(u^{k}_{\text{tr}}\right) \leq 0 , \quad  \ \ \ \ \ \ \quad \forall k = 0, \cdots ,T-1 \ \ \ \forall i = 1,\cdots,N
\end{align}
\end{subequations}

where $T$ is the time horizon, $x_{\text{i}}$, and $u_{\text{i}}$ are the states, and operational set points of hub $i$, respectively. $u_{\text{tr}}$ is the globally shared decision vector for the energy traded between hubs and  $r_{\text{i}}$  and  $s_{\text{i}}$ are selection vectors that only adds the elements of the vector $u_{\text{tr}}$ associated with electrical and thermal energy trades of hub $i$, respectively. The electrical and thermal load demand are $l^{k}_{\text{e,i}}$ and $l^{k}_{\text{h,i}}$ respectively, and $\epsilon^{k}_{\text{i}}$ is the slack variable to ensure a feasible solution. 
The cost function, $ J^{k}_{\text{i}}$, includes the the total cost of importing and exporting electricity from/to the electricity grid, and buying gas from the grid. It also accounts for the fees collected by the network operator for using the grid infrastructure to exchange electrical energy between the hubs; we assume that this cost is borne by the entity importing the energy. Finally, the cost also has the slack variable that is quadratically penalised and weighted by the parameter $\lambda_{\epsilon}$.

The dynamics of the electrical and thermal storages are described by discrete time dynamical systems with a scalar state modelling the state of charge in \eqref{ehub_dyn}. The inequality and equality constraints in \eqref{ehub_eq}, describe the models of the conversion and generation units within the energy hubs as well as their capacity constraints. The load balance equations for the electrical and thermal demands for each hub $i$ are given by \eqref{ehub_loadbalance1}. The net electricity produced in a hub is $e_{\text{i}}(u^{k}_{\text{i}})$ which includes the energy exchanged with the electricity grid and storage devices. The electricity grid acts as a slack in this case and balances any deficit or excess electricity in a hub. Similarly, net heat produced in a hub is $h_{\text{i}}(u^{k}_{\text{i}})$. The total electrical and thermal energy traded by hub $i$ with the other hubs are $r^{\text{T}}_{\text{i}} u^{k}_{\text{tr}}$ and $s^{\text{T}}_{\text{i}} u^{k}_{\text{tr}}$, respectively. These values are positive if energy is imported into a hub and negative otherwise. In the absence of a heating grid, an additional slack is imposed on the thermal load balance equation to ensure feasibility of the optimization problem. The slack represents any unfulfilled thermal demand or any excess heat produced in a hub that is discarded.  The constraints \eqref{ehub_tr1} limit the trade between hubs. Specific trading network topologies can be defined by restricting some of the trading limits to zero.



Energy hubs in a network can be controlled either with a central controller  by solving \eqref{main_opt} or as isolated entities without any trade with other hubs in the network. While central control is proven to provide a globally optimal solution in terms of cost, the resulting optimization problem is large and difficult to solve. To resolve this, \eqref{main_opt} is solved using a distributed algorithm based on the consensus version of the alternating direction method of multipliers (ADMM) wherein the hubs have to reach agreement on the global decision variable. The resulting algorithm and its detailed derivation are provided in~\cite{behrunani_journal}.
\subsection{Multi-horizon MPC}
The optimal control problem is implemented using model predictive control. In this study, a multi-horizon MPC strategy is used in order to increase the horizon without increasing the computational burden or the sampling time for the controller~\cite{behrunani_journal}. The controller operates at multiple horizons, each with a different time resolution. The sampling time is small in the beginning which becomes larger as we move forward in time and the resulting time grid becomes more sparse as we move forward in the horizon. This strategy relies on the property that the sensitivity of the current solution to perturbations decays exponentially as one moves away from the perturbation point and therefore, disruptions in the far future have a negligible impact on the current time. This allows us to have a larger time resolution in the future while maintaining the accuracy of the solution at the first time step which is the only one applied on the system~\cite{diff_mpc}. 


\section{Case Study}

\subsection{Experimental Setup}

\begin{figure}
\centering
  \includegraphics[width=\textwidth]{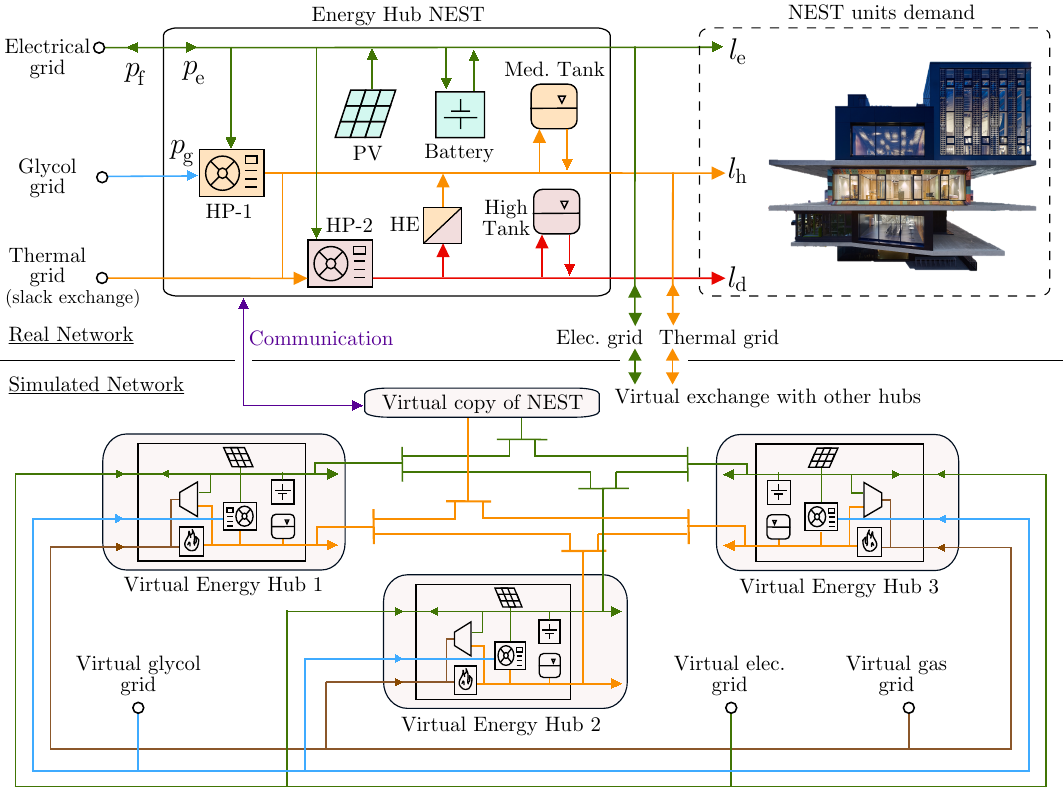}
  \vspace{-0.5cm}
  \caption{Topology of the NEST energy hub, the simulated energy hub network and the interconnection between the real and virtually simulated network. It shows the electrical (green), medium temperature (orange), high temperature (red), glycol (blue), gas (brown) and communication (purple) network in the real and simulated environment.}
  \label{fig:setup}
  \vspace{-0.5cm}
 \end{figure}

The NEST demonstrator consists of an energy hub and several residential and commercial units that serve as a demand for the hub. The energy hub supplies the electrical and thermal demand of 5 units; two residential units, a working space, a fitness centre, and the backbone which houses all units and the hub, which together emulate a small district. In addition to these units, NEST also consists of other units that are not included in this study. Figure \ref{fig:setup} illustrates the topology of the energy hub. The parameters, capacities and detailed model of the NEST hub are available in ~\cite{behrunani_extension}. 
The medium temperature grid supplies the main thermal demand of the building required for space heating and the high temperature grid supplies for domestic hot water. The hub is connected to the electricity grid and glycol grid on the input side. It consists of rooftop and facade photovoltaic (PV), a medium temperature heat pump (HP-1) that draws power from the glycol and electrical grid and injects energy into the medium temperature grid, a high temperature heat pump (HP-2) that uses electricity to convert input from the medium temperature into high temperature, and a heat exchanger between the medium and high temperature grids. The hub also has a battery system for electrical energy storage and storage water tanks for medium and high temperature thermal storage. NEST is also connected to a district medium temperature grid which acts as a slack for the thermal energy. Peer to peer trading between NEST and the virtual hubs is realised using the medium temperature and electricity grids. NEST exports to or imports from the grid to imitate the exchange of energy with the other hubs.



\subsection{Network Topologies}

The NEST district is connected to three virtual energy hubs of different sizes and configurations in each experiment creating a hybrid physical-simulation environment to test the performance of the controller as shown in Figure \ref{fig:setup}. In Experiment 1, we use benchmark energy hubs from the literature to configure a suitable network topology that maximises trading and utility of resources.  Each hub is modelled using its conversion and storage devices, and an overview of the models and constraints used for each component is in \cite{behrunani_stoch, behrunani_journal}. The prices for using energy from the grid and the fees for using the grid for peer to peer trading are obtained from \cite{behrunani_journal, baldini}. The configuration, parameters and capacities for the three virtual hubs, and the prices used in this study are available in~\cite{behrunani_extension}.
In the second experiment, the network includes three buildings in the vicinity of NEST on the Empa campus with their existing technology. The hubs each have rooftop PV and an electrical and thermal demand. The hubs are connected to the Empa medium temperature thermal grid and can import energy at a fixed cost to supply their thermal demand. 

\subsection{Demand and PV Forecasting}

The optimization \eqref{main_opt} requires a forecast of the electrical and thermal demand,  $l^{k}_{\text{e,i}}$ and $l^{k}_{\text{h,i}}$ for each hub $i$ in the network. The aggregated electrical demand of NEST is forecasted using a Gaussian process (GP) based demand predictor from \cite{behrunani_stoch}. A $\SI{72}{\hour}$ ahead demand trajectory is generated by iteratively sampling the one-step GP predictor with a sampling time of $\SI{1}{\hour}$. The high temperature demand of NEST is also forecasted using a similar GP predictor. Finally, the medium temperature thermal demand is predicted using a single ANN proposed in \cite{bunning:2020} to make $\SI{72}{\hour}$ ahead forecasts at a sampling rate of $\SI{1}{\hour}$. The network has one output and the forecast is performed with 72 different input vectors to generate the complete trajectory. The models are trained using all available data from February to May from 2018, 2019 and 2022. We assume that demands of all the other hubs in the network are known perfectly. In Experiment 1, the demands are extracted from the available data set and for Experiment 2, historical demand data for the buildings is used. 

The forecasts for the ambient temperature and global solar radiation are acquired from NEST and are updated once every $\SI{12}{\hour}$. These are used in the demand prediction models as well as for forecasting the PV production. The solar radiation incident to the panel is computed using the forecast of the global radiation and the fixed azimuth and elevation angles using the pvlib package in python and a fixed efficiency is estimated using linear regression on historical output data. The PV output of the hubs in Experiment 1 is estimated in a similar manner as the parameters are known perfectly and the PV output of the buildings in Experiment 2 is extracted from historical data from days that have similar weather conditions.





\section{Results}
\begin{figure}
\centering
  \includegraphics[width=\textwidth]{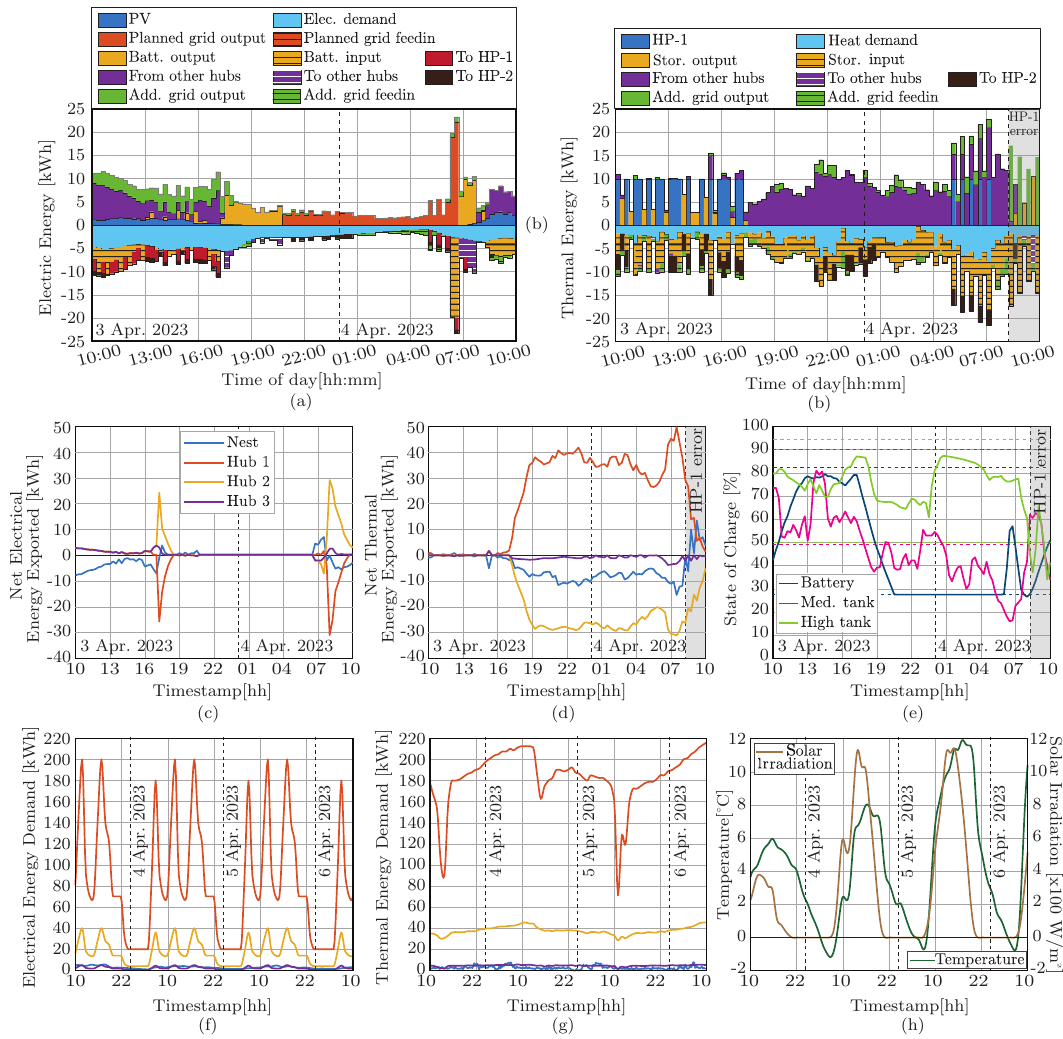}
  \vspace{-0.5cm}
  \caption{Results of Experiment 1.}
  \label{fig:exp1}
  \vspace{-0.5cm}
 \end{figure}

In Experiment 1, the performance of MH-DMPC is evaluated on the energy hub network operation for a period of 3 days, from 10:00 on 3 Apr. 2023 to 10:00 on 6 Apr. 2023. The results are depicted in Fig.~\ref{fig:exp1}. Fig.~\ref{fig:exp1} (a) and (b) shows the dispatch of the different components within the NEST hub for the first $\SI{24}{\hour}$ determined by the controller for electrical and thermal energy respectively. The positive axis shows the energy sources such as PV (for electricity) and HP-1 (for heat) and the negative axis shows the sinks such as energy demands and input to storage devices. Electricity from various sources and energy imported from other hubs is used to fulfil the electricity demand and supply energy to HP-1 and HP-2 to supply the heating demand. Similarly, for heating demand, heat produced by HP-1, discharged from storage and imported from other hubs is used to supply the heating demand and HP-2. The optimization at every time step was performed for a forecast of the energy demands and PV production, and the real values genrally differ from the projected values. This mismatch is balanced by importing additional energy from the grid or feeding excess energy into the grid (green) to balance the supply and demand of each energy carrier and ensure that the electricity and heating demand are always fulfilled in the hub. The net electrical and thermal energy exported by all hubs in the network is shown in Fig.~\ref{fig:exp1} (c) and (d), respectively. NEST mainly imports energy from the other larger hubs in the network and the net energy exported by NEST is negative. Fig.~\ref{fig:exp1} (e) shows the state of charge of the storage devices over this period along with the corresponding maximum and minimum limits imposed. The storage devices charge during the day when PV production is available and discharge at other times. The medium temperature storage tank is also used to supply heat to other units in the NEST that are not included in this experiment which results in the SOC levels often falling below the set minimum. Fig. \ref{fig:exp1} (f) shows the shows the temperature and solar radiation observed during the experiment and the electrical and thermal demands for all the hubs over the span of ${72}$ hours are shown in Fig. \ref{fig:exp1} (g) and (h), respectively. It shows the relative sizes of the other hubs in the network compared to NEST. Hub 1 represents a larger industrial hub with a high production capacity, Hub 2 is a medium sized hub, and Hub 3 is a small residential hub with demand similar to NEST. At the end of day 1, there was an error in HP-1 for some hours (indicated in grey) and this data is excluded in out calculations.

 The second experiment was conducted for period of 2 days, from 18:00 on 6 Apr. 2023 to 18:00 on 8 Apr. 2023 and the results are depicted in \cite{behrunani_extension} (omitted here due to space). The hubs in this case are more comparable in size in terms of their electrical and thermal demands. The NEST demand in this case is also much lower than in the previous experiment due to holidays (7 Apr. and 8 Apr.) and the higher ambient temperature and solar radiation. The lower thermal demand and the increase in ambient temperature also results in the storage tanks being consistently charged.
 
 The total cost of operating the system using the MH-DMPC and the total electricity imported from the grid during the experiments are given in Table \ref{tab:results} and compared to the islanded operation when the hubs in the network are operated without any energy trade under the same conditions and demands. For both experiments, the total cost and energy imports are significantly reduced by leveraging energy trades and efficiently using the available storage resources.

\begin{table}
        \centering
    \begin{tabular}{l c c  c c}
                & \multicolumn{2}{c}{\textbf{Multi-horizon DMPC}} & \multicolumn{2}{c}{\textbf{Islanded operation}}\\ \hline
                  & \multirow{2}{*}{Total cost (CHF)} & Total elec.  &  \multirow{2}{*}{Total cost (CHF)} & Total elec. \\ 
                 & & from grid (kWh) & & from grid (kWh)\\  \hline
                Exp. 1 & -871 & -5,517 & -374 & -5,396\\
                Exp. 2 & 434 & 1,206 & 645 & 1,273 
        \end{tabular}
        \caption{Comparison of the total cost and energy imports from the electricity grid for the complete network using the MH-DMPC and in islanded operation}
                \label{tab:results}  
 \vspace{-0.5cm}
\end{table}

\section{Conclusions}
In this paper, we apply the MH-DMPC control strategy on the real NEST energy hub augmented with a simulated energy hub network to demonstrate the performance and efficiency of the control approach. MH-DMPC has a superior performance in terms of cost and energy consumption in two experiments. The studies show an improved performance both with hubs of different sizes and even within the current building setup where the controller can be immediately implemented. Future studies aim for a large scale implementation of MH-DMPC on real hubs. 
\vspace{-0.3cm}
\section*{Acknowledgments}
This research is supported by the SNSF through NCCR Automation (Grant Number 180545). We also thank Sascha Stoller and Reto Fricker for their technical support.  

\vspace{-0.3cm}
\section*{References}

\bibliography{iopart-num}

\providecommand{\newblock}{}
\begin{thebibliography}{1}
\expandafter\ifx\csname url\endcsname\relax
  \def\url#1{{\tt #1}}\fi
\expandafter\ifx\csname urlprefix\endcsname\relax\def\urlprefix{URL }\fi
\providecommand{\eprint}[2][]{\url{#2}}

\bibitem{MOHAMMADI201833}
Mohammadi M, Noorollahi Y, Mohammadi-ivatloo B, Hosseinzadeh M, Yousefi H and
  Khorasani S~T 2018 {\em Renewable and Sustainable Energy Reviews\/} {\bf 89}
  33--50

\bibitem{behrunani_journal}
Behrunani V, Cai H, Heer P, Smith R and Lygeros J 2023 {\em arXiv preprint
  arXiv:2304.14089\/}

\bibitem{AR_smith}
Smith R~S, Behrunani V and Lygeros J 2023 {\em Annual Review of Control,
  Robotics, \& Autonomous Systems\/} {\bf 6}

\bibitem{NEST}
Richner P, Heer P, Largo R, Marchesi E and Zimmermann M 2017 {\em Informes de
  la Construcción\/} {\bf 69} e222

\bibitem{diff_mpc}
Shin S and Zavala V~M 2023 {\em IEEE Transactions on Automatic Control\/} {\bf
  68} 188--201

\bibitem{behrunani_extension}
Behrunani V 2023 Experimental validation for distributed control of energy hubs
  \url{https://gitlab.ethz.ch/bvarsha/experimental-validation-of-distributed-control-of-energy-hubs}

\bibitem{behrunani_stoch}
Micheli F, Behrunani V, Mehr J, Heer P and Lygeros J 2023 {\em arXiv preprint
  arXiv:2304.12438\/}

\bibitem{baldini}
Baldini L 2016 {\em Sustainable Built Environment (SBE) regional conference,
  Z{\"u}rich\/}

\bibitem{bunning:2020}
Bünning F, Heer P, Smith R~S and Lygeros J 2020 {\em Energy and Buildings\/}
  {\bf 211} 109821

\end{thebibliography}

\end{document}